# Designing Adaptive Robust Extended Kalman Filter Based on Lyapunov-Based Controller for Robotics Manipulators

A.R. Ghiasi[*], A.A. Ghavifekr[**], Y. Shabbouei Hagh[***], H. SeyedGholami[****]

* Department of Electrical and Computer Engineering, University of Tabriz, Tabriz, Iran
Email: agiasi@tabrizu.ac.ir
** Department of Electrical and Computer Engineering, University of Tabriz, Tabriz, Iran
Email: aa.ghavifekr@tabrizu.ac.ir
*** Department of Electrical and Computer Engineering, University of Tabriz, Tabriz, Iran
Email: y.shabbouei92@ms.tabrizu.ac.ir
**** Department of Electrical and Computer Engineering, University of Tabriz, Tabriz, Iran
Email: h.seyedgholami@gmail.com

*Abstract:* In this paper, a position and velocity estimation method for robotic manipulators which are affected by constant bounded disturbances is considered. The tracking control problem is formulated as a disturbance rejection problem, with all the unknown parameters and dynamic uncertainties lumped into disturbance. Using adaptive robust extended Kalman filter(AREKF) the movement and velocity of each joint is predicted to use in discontinuous Lyapunov-based controller structure. The parameters of the error dynamics have been validated off-line by real data.
Computer simulation results given for a two degree of freedom manipulator demonstrate the efficacy of the improved Kalman Filter by comparing the performance of EKF and improved AREKF. Although it is shown that accurate trajectory tracking can be achieved by using the proposed controller.

## 1. Introduction

The tracking control of robotic manipulators has been extensively studied in recent years. The design of high-performance robotic control systems, involving nonlinear control algorithms for robotic mechanisms is of much interest[1]. Various Controllers are designed for manipulators that are subjected to disturbances and parametric uncertainties [2]. In [3] a novel robust passive control approach is introduced for a robot manipulator with model uncertainties to interact with its dynamic environment by considering the robot's mechanic energy. An adaptive position/force controller has been used for robotic manipulator to compensate parametric uncertainties while only requiring the measurements of the position and velocity of robot arms, but not the measurements of forces at contact points[4].Neural networks and fuzzy controllers have been used recently to reduce tracking errors of manipulators[5, 6]. Sliding Mode control approach based on neural network is considered in [7] where the model parameters are used in the equivalent control law. The inertia matrix, the parameters of the matrix of the Coriolis/centripetal terms and the gravity vector are estimated in this method.

There are several inherent difficulties associated with internal and external disturbances. In [8] an adaptive control of a robotic manipulator and a parameter estimation method are discussed. It considers a five-bar linkage manipulator with unknown external disturbance. A novel entropy based indirect iterative learning control methodology for robotic manipulator with random disturbance is sated in [9] by combining the minimum error entropy and the optimal strategy, which is used to update local controller parameters between any two adjacent batches. In [10] disturbance observers (DOB) are used to reject both external disturbances as well as inherent internal uncertainties. As manipulator have to be robust against internal and external variations in the work place the disturbance observer has to estimate unknown states and rejecting decoupled terms between links[11].

In this paper discontinuous Lyapunov-based method is evaluated to reject bounded and time variant disturbances. All system uncertainties are lumped into the disturbance term.

Kalman filters have been used in robotics literature to estimate the joint angles[12, 13]. The main idea behind the EKF is to linearize the model before implementing the Kalman filter. For linearization purpose, it is common to use Taylor expansions. The system can be extended to any orders, but to save the computing resources and also to make the calculation as simple as possible, usually only first (Jacobian) order of Taylor expansion is used. The kalman filter formulation completely depends on a priori knowledge of the system dynamics and it is also assumed that the process and measurement noise covariance matrices are pure white, zero-mean, completely uncorrelated and completely known. If any of these assumptions were violated in any way, as they always were in real implementations, the filter performance will go bad and in some cases may even go divergent. So it can be expected that an adaptive robust extended kalman filter will prevent filter divergence.

The rest of this paper is organized as follows: In section 2 the preliminaries of the problem are established. In section 3 designing kalman filter is formulated into a proper disturbance rejection problem. Discontinuous Lyapunov based controller is proposed in section 4. As an application example, a two degree of freedom manipulator is discussed in 5 and simulation results are presented, and finally the conclusion is given in section 6.

## 2. Preliminaries

Energy-based methods and the methods based on Newton formulation are two general dynamics modeling of robots. Denavit Hartenberg[14] parameters can be used to represent the position of end-effector in Cartesian coordinate relative to ground .
Equation. 1 shows the dynamic model of the manipulator:

$$D(q)\ddot{q} + C(q,\dot{q})\dot{q} + g(q) = u \qquad (1)$$

where $D(q) \in R^{n \times n}$ is the inertia matrix, $C(q,\dot{q})\dot{q} \in R^n$ is the centripetal and Coriolis matrix, $g(q) \in R^n$ is the gravitational force and u is the exerted joint input. $q \in R^n$ is the joint angle vector. Even if the equations of motion of the robot are complex and highly nonlinear, there are still some basic properties in Eq. (1) that are convenient for controller design. These properties are as follows.

**Property 1.** The inertia matrix $D(q)$ is uniformly positive definite, and there exist positive constants $\alpha_1$, $\alpha_2$ such that:

$$\alpha_1 I \leq D(q) \leq \alpha_2 I \qquad (2)$$

**Property 2.** The manipulator dynamics (1) is linear in a set of physical parameters $\theta_d = (\theta_{d1}, \theta_{d2}, ..., \theta_{dp})^T$.

$$D(q)\ddot{q} + C(q,\dot{q})\dot{q} + g(q) = Y(q,\dot{q},\ddot{q})\theta_d \qquad (3)$$

Where $Y(q,\dot{q},\ddot{q}) \in R^{n \times p}$ is usually called the dynamic regressor matrix.

The generalized end-effector's position $x \in R^n$ can be expressed as $x = h(q)$ where $h(.) \in R^n \to R^n$ is generally a nonlinear mapping between joint space and task space

Consider that $d$ is the constant bounded vector describing the load torque, the motion equation of the robot can be written as

$$D(q)\ddot{q} + C(q,\dot{q})\dot{q} + G(q) = u + d \qquad (4)$$

$$\exists\, 0 \leq \delta < \infty;\ \sup_{0 \leq t < \infty} \|d(t)\|_2 < \delta \qquad (5)$$

## 3. Designing Kalman Filters

### 3.1 Extended Kalman Filter

The most vital requirement of the Kalman filter is the linearity of the system. However, in real world, finding linear systems are quite hard and most of the systems that are being dialed with are highly nonlinear, so the Kalman filter should be outstretched that it could handle nonlinear systems too. One of the methods that are being used to overcome this problem is through a linearization procedure. The resulting filter is referred as the extended Kalman filter (EKF).

A non-linear system can be typically formulated as

$$x_k = f(x_{k-1}, u_{k-1}) + w_{k-1} \qquad (6)$$
$$y_k = g(x_k) + r_k \qquad (7)$$

Where $x_k \in R^n$ is state, $y_k \in R^m$ is measurement, $w_k \sim N(0, Q_k)$ is Gaussian noise process, $r_k \sim N(0, R_k)$ is measurement noise, $f$ is non-linear dynamic model function and $g$ is the non-linear measurement model function.

The extended Kalman filter can be summarized into 4 main steps.

1. The state-space model of system are given as follows:

$$x_k = f(x_{k-1}, u_{k-1}) + w_k \qquad (8)$$

$$y_k = g(x_k) + r_k$$

2. Initialize the filter as follows:

$$\hat{x}_0^+ = E[x_0]$$
$$\hat{P}_0^+ = E[(x_0 - E[x_0])(x_0 - E[x_0])^T] \qquad (9)$$

3. For $k = 1, 2, ...$, perform the following:

   3.1. Calculate the Jacobian matrices of $f$ and $h$:

$$F_{k-1} = \frac{\partial f_{k-1}}{\partial x}\Big|_{\hat{x}_{k-1}^+} \qquad (10)$$

$$H_k = \frac{\partial g_k}{\partial x}\Big|_{\hat{x}_k^-}$$

3.2. Perform the time update of the sate estimation and error covariance matrix

$$P_k^- = F_{k-1}P_{k-1}^+ F_{k-1}^T + Q_{k-1} \quad (11)$$
$$\hat{x}_k^- = f_{k-1}(\hat{x}_{k-1}^+, u_{k-1})$$

4. Perform the measurement update:

$$K_k = P_k^- H_k^T (H_k P_k^- H_k^T + R_k)^{-1}$$
$$\hat{x}_k^+ = \hat{x}_k^- + K_k (y_k - h_k(\hat{x}_k^-)) \quad (12)$$
$$P_k^+ = (I - K_k H_k) P_k^-$$

### 3.2 Adaptive Robust Extended Kalman Filter

The AREKF algorithm can be summarized as follows:

1. The nonlinear discrete-time system is represented by

$$x_t = f(x_{t-1}, u_{t-1}) + w_{t-1} \quad (13)$$
$$y_t = g(x_t) + v_t$$

$w_t \sim N(0, Q_t)$ is Gaussian noise process, $v_t \sim N(0, R_t)$ is measurement noise.

2. For $k = 1, 2, \ldots$, perform the following:

2.1 Calculate the Jacobian matrices of $F$ and $G$:

$$F_t = \frac{\partial f_t}{\partial x}\bigg|_{\hat{x}_{t-1}^+} \quad (14)$$
$$G_t = \frac{\partial g_t}{\partial x}\bigg|_{\hat{x}_t^-}$$

2.2 Perform the time update of the sate estimation and error covariance matrix

$$\hat{x}_{t|t-1} = f_{t-1}(\hat{x}_{t-1}, u_{t-1})$$
$$P_{t|t-1} = F_t P_{t-1} F_t^T + Q_t$$

$$\Xi_{t|t-1} = \begin{cases} P_{t|t-1}, & P_{y,t} > \alpha \bar{P}_{y,t} \\ (P_{t|t-1}^{-1} - \gamma^{-2}\Gamma_t^T \Gamma_t)^{-1}, & \text{otherwise} \end{cases}$$

$$\Gamma_t = \gamma(P_{t-1}^{-1} - \lambda_t^{-1} P_{t-1}^-)^{1/2} \quad (15)$$

$$\bar{P}_{y,t} \approx \begin{cases} \tilde{y}_t \tilde{y}_t^T, & t = 0 \\ \dfrac{\rho \bar{P}_{y,t-1} + \tilde{y}_t \tilde{y}_t^T}{\rho + 1}, & t > 0 \end{cases}$$

Where $\bar{P}_{y,t} = E(\tilde{y}_t \tilde{y}_t^T | \tilde{x}_{t-1})$ is the real covariance matrix of the $\tilde{y}_t = y_t - G_t \hat{x}_{t|t-1}$. $\alpha > 0$ is an extra tuning parameter that gives more degree of freedom. $\rho = .98$ is a forgetting factor.

The tuning parameter $\gamma$ is found by searching over $\gamma \neq 0$ and $\lambda_t$ is another tuning parameter which should be large enough that $\Xi_{t|t-1} < \lambda_t P_{t|t-1}$ is fulfilled.

3. Perform the measurement update:

$$\hat{x}_t = \hat{x}_{t|t-1} + K_t (y_t - g_t(\hat{x}_{t|t-1}))$$
$$K_t = \Xi_{t|t-1} G_t^T P_{y,t}^{-1}$$
$$P_{y,t} = G_t \Xi_{t|t-1} G_t^T + R_t \quad (16)$$
$$P_t = (\Xi_{t|t-1}^{-1} + G_t^T R_t^{-1} G_t)^{-1}$$

### 4. Manipulator Control Law Formulation

The proposed discontinuous Lyapunov-based method can eliminate time-varying and bounded disturbances. The proposed controller is:

$$u = D(q)\ddot{\zeta} + C(q,\dot{q})\dot{\zeta} + G(q) - K_D \frac{\sigma}{\|\sigma\|_2} \quad (17)$$

where $K_D = k_d \times I_{n \times n}$ is the positive diagonal matrix and $\sigma = \dot{q} - \dot{\zeta}$. By applying the above control law, closed loop form of equation can be written as following:

$$D(q)\dot{\sigma} + C(q,\dot{q})\sigma + K_D \frac{\sigma}{\|\sigma\|_2} = d \quad (18)$$

After applying controller closed-loop dynamic will be globally asymptotically stable, for which the proof is given in theorem.1. It should be mentioned that all states estimated by proposed Kalman filters.

**Theorem 1.** Choose proposed discontinuous Lyapunov based control law as (17). According to (4) and (5) in the presence of tremor, the closed-loop dynamic that obtained as (18) is asymptotically globally stable if $k_d > \delta$.

**Proof.** Let consider the Lyapunov function candidate as

$$V := \frac{1}{2}\sigma^T D(\hat{q})\sigma \Rightarrow \dot{V} = \sigma^T D(q)\dot{\sigma} + \frac{1}{2}\sigma^T \dot{D}(\hat{q})\sigma \quad (19)$$

Differentiating $V$ with respect to time leads to

$$\dot{V} = \sigma^T \left(-C(\hat{q},\dot{\hat{q}})\sigma - K_D \frac{\sigma}{\|\sigma\|_2} + d\right) + \frac{1}{2}\sigma^T \dot{D}(\hat{q})\sigma$$

$$\dot{V} = \frac{1}{2}\underbrace{\sigma^T \left(\dot{D}(\hat{q}) - 2C(\hat{q},\dot{\hat{q}})\right)\sigma}_{=0} - K_D \|\sigma\|_2 + \sigma^T d \quad (20)$$

$$\dot{V} \leq -K_D \|\sigma\|_2 + \|\sigma\|_2 \delta = -(K_D - \delta I_{n \times n})\|\sigma\|_2$$

Assumption of $k_d > \delta$ and $\hat{k}_d := k_d - \delta$

$$\dot{V} \leq -\hat{k}_d \|\sigma\|_2 \quad (21)$$

According to quadratic functions' properties

$$\frac{1}{2}\lambda_m \|\sigma\|_2^2 \leq V = \frac{1}{2}\sigma^T D(q)\sigma \leq \frac{1}{2}\lambda_M \|\sigma\|_2^2 \qquad (22)$$

Combination of (20) and (21) leads to

$$\dot{V} \leq -\hat{k}_d \sqrt{\frac{2V}{\lambda_M}} = -\alpha\sqrt{V}, \quad \alpha := \hat{k}_d \sqrt{\frac{2}{\lambda_M}} > 0 \qquad (23)$$

Using comparison lemma

$$V(t) \leq \left(\sqrt{V(0)} - \frac{\alpha}{2}t\right)^2 \Rightarrow$$

$$\frac{1}{2}\lambda_m \|\sigma(t)\|_2^2 \leq V(t) \leq \left(\sqrt{V(0)} - \frac{\alpha}{2}t\right)^2$$

$$\frac{1}{2}\lambda_m \|\sigma(t)\|_2^2 \leq \left(\sqrt{\frac{\lambda_M}{2}}\|\sigma(0)\| - \frac{\alpha}{2}t\right)^2 \Rightarrow \qquad (24)$$

$$\|\sigma(t)\|_2 \leq \sqrt{\frac{2}{\lambda_m}} \left|\sqrt{\frac{\lambda_M}{2}}\|\sigma(0)\| - \frac{\alpha}{2}t\right|$$

Since t is a positive parameter,

$$0 \leq t_{set} \leq \frac{\sqrt{2\lambda_M}}{\alpha}\|\sigma(0)\| \qquad (25)$$

Where $t_{set}$ is the maximum time takes $\sigma \to 0$.

## 5. Simulation

In this part the simulation results of declared methods have been presented. The planer elbow of manipulator that illustrated in Fig. 1 is considered with the parameters that are given in Table I.

TABLE I
PHYSICAL PARAMETERS OF THE PLANER ELBOW MANIPULATOR

| ith body | $m_i(Kg)$ | $I_i(Kgm^2)$ | $l_i(m)$ | $l_{ci}(m)$ |
|---|---|---|---|---|
| 1 | 1.00 | 0.25 | 0.5 | 0.25 |
| 2 | 1.00 | 0.25 | 0.5 | 0.25 |

The sample time is set to Ts=.005, the true process and measurement noise covariances that effects the model are $10^{-5}I_{4*4}$ nd $10^{-3}I_{4*4}$ respectively, but it is assumed that the noise covariances that are needed in the algorithmare set to be $10^{-9}I_{4*4}$ nd $10^{-1}I_{4*4}$ or process and measurement noise. On the other hand, to show the effectiveness of AREKF than EKF, it is supposed that there are also model uncertainties on some parameters. The tuning parameters that are needed for AREKF can be found on Table II.

TABLE II
TUNING PARAMETERS USED IN AREKF

| ρ | α | λ | γ |
|---|---|---|---|
| .97 | .9 | .7 | .001 |

The control goal is tracking desired trajectory in joint space in the presence of constant bounded disturbances. Disturbance modeling leads to terms that are added with the input. Closed-form expressions for the mass-inertia matrix $D(q)$, the Coriolis and centrifugal matrix $C(q,\dot{q})$, and the gravity vector $G(q)$ are obtained.

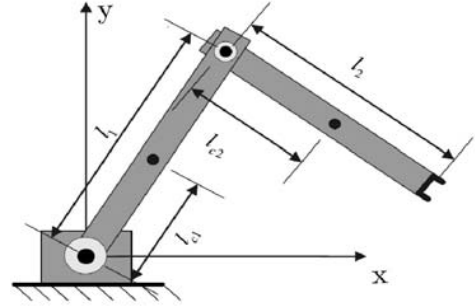

Fig. 1. A two-DOF manipulator

Appropriate control law can be written as:

$$\begin{cases} u = D(q)\ddot{\zeta} + C(q,\dot{q})\dot{\zeta} + G(q) - K_D \frac{\sigma}{\|\sigma\|_2} \\ \sigma = \dot{\tilde{q}} + \Lambda\tilde{q}, \quad \dot{\xi} = \dot{q}_d - \Lambda\tilde{q} \end{cases} \qquad (26)$$

Where the matrices $K_D, \Lambda$ are symmetric positive definite. To design the controller, gain matrices are assumed to be $K_D = 9 \times I_2, \Lambda = 3 \times I_2$.

Figures 2-5 represent the position and velocity of first and second joint estimations. Control signal and trajectory tracking errors beside $\sigma$ variations with respect to time, are presented in figures 6 and 7 respectively.

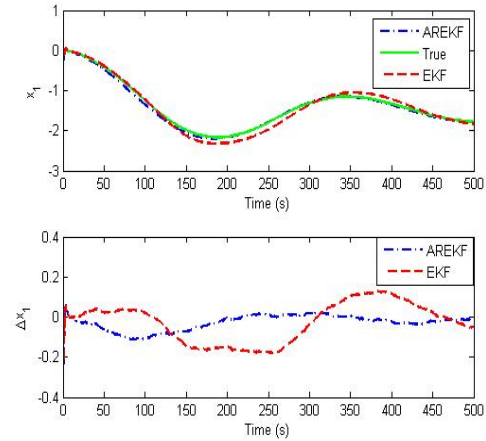

Fig.2 Comparison of estimated states of 2-DOF robot: the position of the fisrt joint

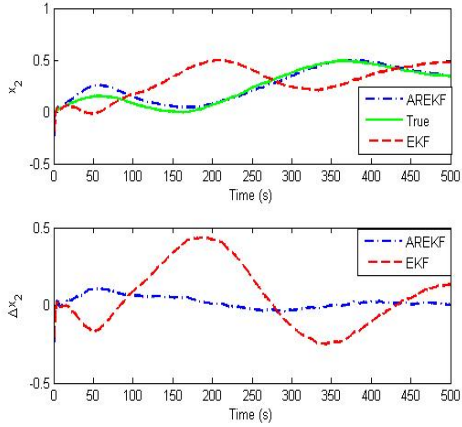

Fig.3 Comparison of estimated states of 2-DOF robot: the velocity of the fisrt joint

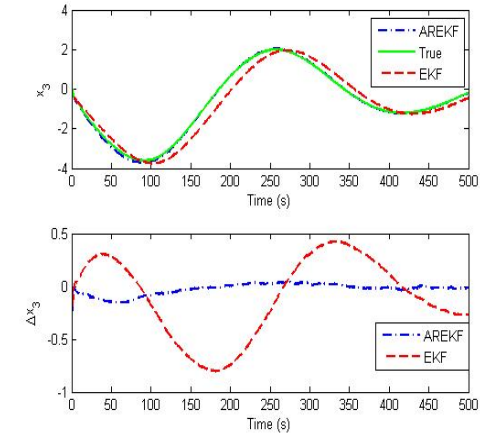

Fig.4 Comparison of estimated states of 2-DOF robot: the position of the second joint

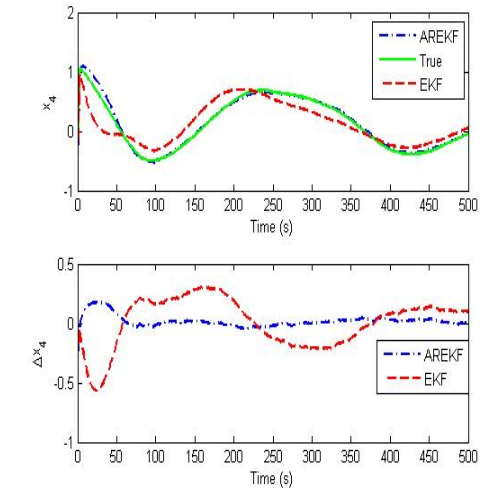

Fig.5 Comparison of estimated states of 2-DOF robot: the velocity of the second joint

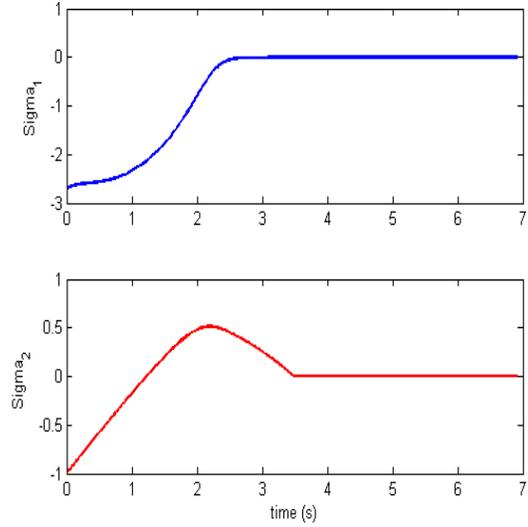

Fig. 6 Evaluation of $\sigma$ with respect to time

The *'sign'* function is a discontinuous function and its value in zero can cause problem by stimulating unmodeled high order dynamics of system. To solve this problem, a saturation function is used instead of 'sign' in control laws.

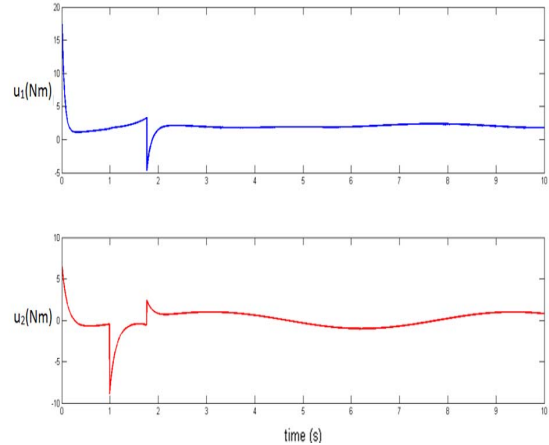

Fig. 7. Control input signals based on saturated controller

## 6. Conclusion

Adaptive robust extended kalman filter based on discontinuous Lyapunov-based controller is presented in this paper to improve the accuracy of tracking procedure for robotic manipulators with time variant disturbances.

The results of application of the AREKF for two degree of freedom manipulator are enough encouraging.
In this case, the Extended Kalman Filter can quickly converge and avoid trapped in local minimum of the function.
The tracking control problem is formulated as a disturbance rejection problem, with all the system nonlinearities and uncertainties lumped into disturbance. Simulation results for two degree of freedom manipulator shows the efficacy of estimation and accurate tracking. The improvement of estimation algorithm and domain of attraction in stability proof can be considered in future work.